# Quantum Cryptography for Enhanced Network Security: A Comprehensive Survey of Research, Developments, and Future Directions

Mst Shapna Akter*

*Department of Computer Science, Kennesaw State University, USA
{* makter2@students.kennesaw.edu}

*Abstract*—With the ever-growing concern for internet security, the field of quantum cryptography emerges as a promising solution for enhancing the security of networking systems. In this paper, 20 notable papers from leading conferences and journals are reviewed and categorized based on their focus on various aspects of quantum cryptography, including key distribution, quantum bit commitment, post-quantum cryptography, and counterfactual quantum key distribution. The paper explores the motivations and challenges of employing quantum cryptography, addressing security and privacy concerns along with existing solutions. Secure key distribution, a critical component in ensuring the confidentiality and integrity of transmitted information over a network, is emphasized in the discussion. The survey examines the potential of quantum cryptography to enable secure key exchange between parties, even when faced with eavesdropping, and other applications of quantum cryptography. Additionally, the paper analyzes the methodologies, findings, and limitations of each reviewed study, pinpointing trends such as the increasing focus on practical implementation of quantum cryptography protocols and the growing interest in post-quantum cryptography research. Furthermore, the survey identifies challenges and open research questions, including the need for more efficient quantum repeater networks, improved security proofs for continuous variable quantum key distribution, and the development of quantum-resistant cryptographic algorithms.

## I. INTRODUCTION

The emergence of quantum computing has brought forth both challenges and opportunities in the realm of cryptography. Quantum computers hold the potential to revolutionize various industries by tackling complex problems; however, they also present a significant threat to existing cryptographic systems' security [1]. Consequently, researchers have turned to quantum cryptography, utilizing quantum mechanics principles to create secure communication systems that withstand both classical and quantum attacks. Quantum key distribution (QKD) is a cryptographic technique that enables two parties to securely exchange encryption keys over a public channel [2]. QKD protocols exploit the fundamental properties of quantum mechanics, such as superposition, entanglement, and the no-cloning theorem, to ensure that any eavesdropping attempt can be detected, thus providing information-theoretic security [3]. In contrast to classical cryptographic techniques, which rely on the computational difficulty of solving certain mathematical problems, QKD guarantees security even against adversaries with unlimited computational power [4]. Over the years, QKD has garnered significant attention from both academia and industry, leading to the development of various QKD protocols, such as BB84 bennett2020quantum, E91 [5], and continuous variable QKD [6]. These protocols have been the subject of extensive research, with efforts dedicated to improving their efficiency, security, and applicability to real-world communication networks [7]. One major area of research in quantum cryptography has been the development and optimization of QKD protocols. Researchers have investigated different approaches to optimize key rates, reduce the quantum bit error rate, and increase the distance over which secure communication can be achieved [8]. These optimizations have led to the proposal of new protocols, such as measurement-device-independent QKD (MDI-QKD) [9] and twin-field QKD (TF-QKD) [10], which offer improved performance and robustness against various types of attacks. In addition to the development of new protocols, researchers have also focused on identifying and mitigating potential security loopholes in existing QKD protocols [11]. For example, photon-number-splitting attacks and detector blinding attacks have been shown to compromise the security of several QKD implementations. Various countermeasures have been proposed and implemented to address these vulnerabilities, such as the decoy-state method [12] and the use of secure detectors [13]. Another crucial aspect of quantum cryptography research is the integration of QKD into existing communication networks. One approach has been to incorporate QKD into optical networks, which form the backbone of modern communication infrastructure [14]. Several studies have investigated the feasibility of implementing QKD in wavelength-division multiplexing (WDM) networks and passive optical networks (PONs)

[15]. These studies demonstrate the potential of QKD to enhance the security of optical networks without significantly affecting their performance. The integration of QKD into optical networks has also led to the development of new service models, such as Key-as-a-Service (KaaS) [8]. KaaS provides secure key distribution for virtual optical networks (VONs) by incorporating QKD into the underlying optical infrastructure. By offering security as a service, KaaS enables network operators to easily deploy QKD-based security solutions in existing networks, potentially paving the way for widespread adoption of quantum cryptography. Moreover, as quantum computing technology progresses, it has become increasingly important to explore cryptographic techniques that can withstand the potential threat posed by quantum computers. This has led to the emergence of post-quantum cryptography, a field dedicated to developing cryptographic algorithms that remain secure even in the presence of quantum adversaries [16, 17]. Lattice-based cryptography, code-based cryptography, and isogeny-based cryptography are among the most promising post-quantum cryptographic techniques being investigated [18]. While quantum cryptography has shown tremendous potential for enhancing network security, several challenges and open research questions remain to be addressed. For instance, the development of efficient quantum repeater networks is essential to increase the range of QKD systems [19]. Improved security proofs for continuous variable QKD and other protocols are necessary to ensure their robustness against potential attacks [6, 7]. Furthermore, the practical implementation of quantum cryptography systems, including miniaturization, cost reduction, and compatibility with existing infrastructure, is a critical area of ongoing research [20].

## II. RELATED WORK

In this literature review, we analyze the advancements and challenges in the field of quantum cryptography, focusing on quantum key distribution (QKD), post-quantum cryptography, and the integration of QKD into optical networks. A total of 20 papers, along with additional related works, were selected from leading conferences and journals, including

**Quantum Key Distribution (QKD) Protocols**

QKD protocols have been extensively studied to enable secure key exchange between two parties. The seminal BB84 protocol, introduced by Bennett and Brassard [2], is one of the earliest and most widely studied QKD protocols. Subsequent research led to the development of other QKD protocols, such as the E91 protocol [5] and continuous variable QKD [6]. Each protocol leverages the unique properties of quantum mechanics to provide information-theoretic security [3].

Nurhadi and Syambas [21] provide an overview of various QKD protocols, including BB84, E91, BBM92, B92, Six-State Protocol, DPS, SARG04, COW, and S13. The authors then conduct simulations of three of these protocols, BB84, B92, and BBM92, using a quantum simulator. The results show that B92 protocol has the smallest probability of error, while BB84 has the largest probability of error. Kalra and Poonia [22] propose a new protocol that is a variation of the BB84 protocol and show that it is twice as capacitive as compared to the BB84 protocol, with almost half the error rate. The proposed protocol uses random bases for modulation and encoding on the basis of random bits, and both the sender and the receiver get two keys. Sasaki et al. [23] propose a QKD protocol that uses a single-photon source to generate a sequence of pulses, each containing one or zero photons, which is sent to a receiver. The security of the protocol relies on the laws of quantum mechanics and the assumption that any measurement or disturbance by an eavesdropper can be detected. Dirks et al. [24] explore the technical feasibility of a Geostationary Earth Orbit Quantum Key Distribution (GEOQKD) system that combines untrusted and trusted mode BBM92 protocols to achieve a maximum tolerable loss of 41dB per channel, with key rates of 1.1bit/s in untrusted and 300bit/s in trusted mode. The study proposes a realistic design for the space segment and presents a system architecture that allows the GEOQKD system to operate in both untrusted and trusted modes with high pointing accuracies. Williams et al. [25] present a QKD protocol that uses time-bin encoding with entangled photon pairs to achieve secure communication. The protocol was implemented in a practical setup and was tested to demonstrate time synchronization and eavesdropper detection capabilities. Schimpf et al. [26] discussed a study on using a blinking-free source of polarization-entangled photon pairs based on a GaAs QD for QKD. The study addresses the problem of degradation of entanglement at higher temperatures and proposes to operate the source at a temperature of at least 20 K and to use a pulsed two-photon-excitation scheme to maintain fidelity to the Bell state. Amer et al. [27] presented a study on the performance of quantum repeater QKD grid networks with the inclusion of a minority of trusted nodes. The analysis also identifies limitations in such networks, particularly related to BSM success probability and decoherence rate, and suggests the use of trusted nodes even with ideal repeater technology. Ding et al. [28] proposed a new approach to optimize the parameters of practical QKD systems using the random forest (RF) algorithm. The proposed method has potential applications in practical QKD networks and contributes to the development of quantum communication technologies. Dhoha et al. [29] provided a literature review of QKD and quantum bit commitment (QBC) protocols. The focus of the paper is on the practical implementation of the BB84 QKD protocol, both with and without the existence of an eavesdropper. The findings show that BB84 is an effective QKD protocol. Yao et al. [30] discuss the use of quantum random number generators (QRNGs and QKD protocols in cryptography, and provide a theoretical analysis of their security based on entropic uncertainty relations. The authors use Theorem II.1 to show that by choosing suitable classical sampling strategies, one may analyze the behavior of ideal states which always behave appropriately for the given strategy, and that the real state is close, in trace distance, to these ideal states.

**Post-Quantum Cryptography**

Mujdei et al. [31] investigated side-channel attacks on Kyber, Saber, and NTRU post-quantum cryptographic schemes. They proposed a new attack strategy and demonstrated its effectiveness against countermeasures like randomization techniques. This study highlights the importance of considering side-channel attacks in post-quantum cryptography design and implementation. Imana et al. [32] proposed two efficient architectures for arithmetic operations in InvBRLWE-based encryption, improving area-time complexities and power efficiency. The authors provided a theoretical analysis and FPGA-based implementation, showing potential for use in BRLWE/InvBRLWE-based cryptoprocessor applications. Prakasan et al. [33] addressed security issues in the classical channel of Quantum Key Distribution (QKD) by proposing an authenticated-encryption scheme using NTRU and Falcon algorithms. The scheme enhances security without significant performance trade-offs and offers a viable solution for QKD security concerns. Sajimon et al. [34] evaluated PQC algorithms for IoT devices and identified Kyber, Saber, Dilithium, and Falcon as optimal implementations. The study also recommended LightSaber-KEM and Dilithium2 for quantum resistance. The research methodology involved using Raspberry Pi 4 for performance evaluation and can be extended to assess quantum-resistant TLS and DTLS schemes for IoT.

**Security Issues and Countermeasures**

Abidin et al. [35] discussed the use of quantum cryptography and QKD in the DARPA Quantum Network for secure VPN communication. The study elaborated on QKD protocols, algorithms, and their implementation with IPsec. The article highlights the promising nature of quantum cryptography for securing cyberspace and addressing internet security concerns. Kumar et al. [36] examined various post-quantum cryptographic approaches for securing IoT networks. The paper compared recent work in this area and concluded that lightweight and secure post-quantum cryptography for small devices is expected to emerge in the near future. Ahn et al. [37] analyzed the potential impact of quantum computing on DER networks and proposed using PQC and QKD to protect them. The study suggested researching optimal cost and network configuration for cost-effective and high-performance quantum-safe networks in DER systems. Gupta et al. [38] explored the use of blockchain technology in e-voting systems and proposed a double-layered security system that uses a QKD algorithm for secure communication. The study highlights the potential for future research in blockchain with quantum computer countermeasures. Lin et al. [39] identified security loopholes in CV-QKD and proposed modifications to existing protocols. The study suggested further research to develop security proofs based on collective attacks and practical source and channel loss. Cao et al. [40] proposed a KaaS framework for integrating QKD into optical networks, enhancing their security. The performance evaluation demonstrated the framework's potential as a practical solution for incorporating QKD in optical networks. Su et al. [41] presented a simple information-theoretic proof of security for the BB84 QKD protocol. The findings provide a clear and straightforward proof of security, offering new insights into security issues in quantum key distribution.

### III. QUANTUM KEY DISTRIBUTION

Quantum Key Distribution (QKD) is a method of secure communication that uses quantum mechanics to distribute cryptographic keys between two parties. The basic idea is that the act of measuring a quantum system disturbs it in a detectable way, so any eavesdropper trying to intercept the key would leave a trace. Alice and Bob generate a shared key by exchanging quantum states (such as photons) and measuring them in a particular way. By comparing their measurements, they can detect any attempted eavesdropping and use the remaining key bits to establish a secret key for encrypting and decrypting messages. QKD offers perfect secrecy, meaning that the encrypted message cannot be deciphered by an eavesdropper, but it has limitations in terms of distance and speed of communication.

### IV. MOTIVATION AND CHALLENGES

The increasing dependence on digital technologies has led to a growing demand for secure and privacy-preserving cryptographic protocols. Quantum cryptography has emerged as a promising solution to address these challenges, particularly in the field of cryptocurrency. Quantum cryptocurrency involves the use of quantum cryptography protocols to provide secure transactions that are resistant to attacks from quantum computers. However, the implementation of these protocols poses several challenges, and security and privacy issues need to be carefully considered. One of the primary challenges in the implementation of quantum cryptocurrency is the development of secure quantum key distribution (QKD) protocols. QKD protocols provide a secure method for generating shared secret keys between two parties that can be used for cryptographic applications. Several QKD protocols have been proposed, including BB84, E91, and B92. However, these protocols are vulnerable to attacks from quantum computers, and more robust protocols need to be developed. Another challenge in the implementation of quantum cryptocurrency is the development of post-quantum cryptographic algorithms. Post-quantum cryptography refers to cryptographic algorithms that are resistant to attacks from both classical and quantum computers. While several post-quantum cryptographic algorithms have been proposed, such as lattice-based cryptography, code-based cryptography, and hash-based cryptography, they are not yet widely adopted, and more research is needed to ensure their security and efficiency. Security and privacy issues also need to be carefully considered in the implementation of quantum cryptocurrency. One of the primary security concerns in quantum cryptocurrency is the possibility of quantum hacking. Quantum hacking involves intercepting and manipulating the qubits used in quantum cryptography protocols, which can compromise the security of the system. Several countermeasures have been proposed to prevent quantum hacking, such as decoy state methods and entanglement-based QKD protocols. Privacy is another important consideration in quantum cryptocurrency. While quantum cryptography protocols provide a high degree of security, they do not necessarily provide privacy. For example, in QKD protocols, the privacy of the communication depends on the ability of the two parties to keep the secret key secure. If one party's system is compromised, the privacy of the communication can be compromised as well. Solutions to these issues include privacy amplification protocols and quantum coin flipping protocols. Several research papers have been published on the topic of quantum cryptocurrency, proposing various solutions to the challenges and issues mentioned above. Table 1 provides an overview of the papers reviewed in this survey, including their focus, methodology, findings, and limitations. The papers cover a range of topics, including quantum key distribution, post-quantum cryptography, counterfactual quantum key distribution, and key management. Through the survey, we aim to provide a comprehensive analysis of the current state of research in quantum cryptocurrency and identify key challenges and future research directions.

| Category | Algorithms/Protocols | Source | Findings | Challenges |
|---|---|---|---|---|
| QKD Protocols | BB84, E91, BBM92, B92, Six-State Protocol, DPS, SARG04, COW, S13 | Nurhadi et al. [21] | B92 has the smallest probability of error | - |
| QKD Protocols | BB84 variation | Kalra and Poonia [22] | Twice as capacitive as BB84 with almost half the error rate | - |
| QKD Protocols | Single-photon source protocol | Sasaki et al. [23] | Secure key distribution based on quantum mechanics | - |
| QKD Protocols | GEOQKD system | Dirks et al. [24] | Achieves maximum tolerable loss of 41dB per channel | - |
| QKD Protocols | Time-bin encoding with entangled photon pairs | Williams et al. [25] | Demonstrates time synchronization and eavesdropper detection | - |
| QKD Protocols | GaAs QD for QKD | Schimpf et al. [26] | Maintains fidelity to the Bell state at higher temperatures | Degradation of entanglement at higher temperatures |
| QKD Protocols | Quantum repeater QKD grid networks | Amer et al. [27] | Identifies limitations in BSM success probability and decoherence rate | - |
| QKD Protocols | Random forest algorithm for QKD parameter optimization | Ding et al. [28] | Contributes to the development of quantum communication technologies | - |
| QKD and QBC Protocols | BB84 | Dhoha et al. [29] | Effective QKD protocol | - |
| QRNG and QKD | Entropic uncertainty relations | Yao et al. [30] | Analyzes behavior of ideal states for QRNG and QKD | - |
| Post-Quantum Cryptography | Kyber, Saber, NTRU | Mujdei et al. [31] | Proposed new attack strategy against countermeasures | Side-channel attacks |
| Post-Quantum Cryptography | InvBRLWE-based encryption | Imana et al. [32] | Improved area-time complexities and power efficiency | - |
| Post-Quantum Cryptography | NTRU and Falcon algorithms | Prakasan et al. [33] | Enhances security without significant performance trade-offs | - |
| Post-Quantum Cryptography | Kyber, Saber, Dilithium, Falcon | Sajimon et al. [34] | Optimal implementations for IoT devices | - |
| Security Issues and Countermeasures | QKD in DARPA Quantum Network | Abidin et al. [35] | Promising nature of quantum cryptography for securing cyberspace | - |
| Security Issues and Countermeasures | Post-quantum cryptographic approaches for IoT | Kumar et al. [36] | Lightweight and secure post-quantum cryptography for small devices is expected to emerge | - |
| Security Issues and Countermeasures | QKD in DER networks | Ahn et al. [37] | Proposes using PQC and QKD to protect DER networks | Optimal cost and network configuration for quantum-safe networks |
| Security Issues and Countermeasures | Blockchain with QKD | Gupta et al. [38] | Proposed double-layered security system using QKD algorithm for secure communication | - |
| Security Issues and Countermeasures | CV-QKD modifications | Lin et al. [39] | Identifies security loopholes in CV-QKD | Security proofs based on collective attacks and practical source/channel loss |
| Security Issues and Countermeasures | Integrating QKD into optical networks | Cao et al. [40] | Proposed KaaS framework for incorporating QKD in optical networks | - |
| Security Issues and Countermeasures | BB84 QKD protocol security proof | Su et al. [41] | Provides a simple information-theoretic proof of security for BB84 | - |

TABLE I: Summary of advancements and challenges in quantum cryptography.

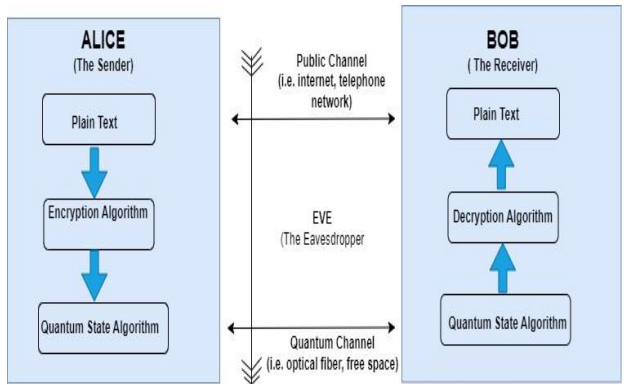

Fig. 1: Basic Block Diagram of QKD System

## V. RESULTS AND DISCUSSION

Our review of the literature on quantum cryptography, including quantum key distribution (QKD), post-quantum cryptography, and their integration into optical networks, has led to several significant findings and highlighted areas for further discussion.

**QKD Protocols:** Various QKD protocols such as BB84, E91, B92, and others have been developed to enable secure key exchange between parties. While each protocol leverages the unique properties of quantum mechanics to provide information-theoretic security, they face challenges in terms of performance, efficiency, and potential vulnerabilities. Further research and optimization of these protocols are required to enhance their practical implementation in quantum communication systems.

**Post-Quantum Cryptography:** Several post-quantum cryptographic techniques, including lattice-based cryptography, code-based cryptography, and isogeny-based cryptography, are being explored to develop cryptographic algorithms that remain secure in the presence of quantum adversaries. These algorithms show promise, but more research is needed to ensure their security, efficiency, and wide adoption in the face of quantum threats.

**Integration of QKD into Optical Networks:** The integration of QKD into optical networks, such as Key-as-a-Service (KaaS) models, has led to the development of new service models and facilitated the deployment of QKD-based security solutions in existing networks. This advancement paves the way for widespread adoption of quantum cryptography. However, practical implementation challenges, including miniaturization, cost reduction, and compatibility with existing infrastructure, remain to be addressed.

**Security Issues and Countermeasures:** Quantum hacking, side-channel attacks, and other vulnerabilities pose challenges to the security of quantum cryptography systems. Countermeasures such as decoy state methods, entanglement-based QKD protocols, privacy amplification protocols, and quantum coin flipping protocols have been proposed to mitigate these threats. Further research is needed to develop robust security measures that can withstand the evolving threat landscape.

**Quantum Cryptocurrency:** The implementation of quantum cryptography in cryptocurrency presents unique challenges, including secure QKD protocols, post-quantum cryptographic algorithms, and privacy concerns. While research has been conducted to address these challenges, more work is needed to develop secure and efficient quantum cryptocurrency systems.

Quantum cryptography holds significant potential for enhancing network security and privacy. Despite the progress made in the field, several challenges and open research questions remain. Addressing these challenges and advancing the state of research in quantum cryptography will contribute to the development of secure communication technologies and pave the way for practical applications, such as quantum cryptocurrency.

## VI. CHALLENGES AND OPEN RESEARCH QUESTIONS

The following challenges and open research questions have been identified based on our review of the literature on quantum cryptography and quantum cryptocurrency:

1. **Robust and Efficient QKD Protocols:** The development of practical, efficient, and robust QKD protocols is crucial for the widespread adoption of quantum cryptography. Further research is needed to optimize existing protocols, address potential vulnerabilities, and devise new protocols that can withstand advanced attacks, including those from quantum adversaries.

2. **Post-Quantum Cryptographic Algorithm Development and Standardization:** As the field of post-quantum cryptography advances, more research is needed to ensure the security, efficiency, and interoperability of post-quantum cryptographic algorithms. Additionally, the development of standardized cryptographic algorithms and protocols that can be widely adopted by industry and government is critical for securing communication systems against quantum threats.

3. **Quantum-Resistant IoT Devices:** With the increasing prevalence of IoT devices, it is essential to develop lightweight and efficient cryptographic solutions that can be implemented on resource-constrained devices. Research should focus on optimizing post-quantum cryptographic algorithms for IoT devices and exploring efficient QKD solutions tailored for IoT environments.

4. **Secure Key Management and Storage:** The security of quantum cryptography systems depends on the secure management and storage of cryptographic keys. Research should explore novel approaches for key management, distribution, and storage that can maintain security even in the presence of quantum threats.

5. **Quantum Cryptocurrency Security and Privacy:** In the context of quantum cryptocurrency, there is a need to address specific security and privacy challenges. Research should focus on the development of secure and private quantum cryptocurrency systems, including the integration of privacy-preserving techniques and novel protocols that can protect user privacy while maintaining the security of transactions.

6. **Scalability and Interoperability:** Practical implementation of quantum cryptography solutions requires scalable and interoperable systems that can seamlessly integrate with existing communication infrastructure. Research should focus on developing scalable quantum cryptography systems and protocols that can be easily deployed and integrated with existing networks and technologies.

7. **Experimental Demonstration and Deployment:** While many quantum cryptography protocols and algorithms have been proposed and analyzed theoretically, there is a need for more experimental demonstrations and real-world deployments. Experimental research should focus on validating and optimizing protocols, algorithms, and countermeasures in realistic settings to better understand their performance and limitations.

8. **Quantum Hacking and Countermeasures:** As quantum computing advances, the potential for quantum hacking and other sophisticated attacks grows. Research should focus on identifying and addressing potential security vulnerabilities in quantum cryptography systems and developing robust countermeasures that can withstand evolving threats.

Addressing these challenges and open research questions will contribute to the development of secure and practical quantum cryptography solutions and pave the way for applications such as quantum cryptocurrency, enhancing the security and privacy of digital communication in the quantum era.

## VII. FUTURE DIRECTIONS

The development and optimization of robust QKD protocols that can withstand advanced attacks, as well as the exploration of secure and efficient post-quantum cryptographic algorithms, ensuring their interoperability and standardization. As IoT devices become more prevalent, lightweight and efficient cryptographic solutions tailored for resource-constrained devices will be crucial. This will involve optimizing post-quantum cryptographic algorithms and QKD solutions for IoT environments. In addition, research should explore novel approaches for key management, distribution, and storage in quantum cryptography systems to maintain security in the face of quantum threats. The development of secure and private quantum cryptocurrency systems is another important area for research, which involves integrating privacy-preserving techniques and novel protocols to protect user privacy while maintaining transaction security. The scalability and interoperability of quantum cryptography systems are essential for practical implementation, so future research should concentrate on creating systems that can be easily deployed and integrated with existing networks and technologies. Experimental demonstrations and real-world deployments of quantum cryptography protocols and algorithms will be critical to validate their performance and limitations. Lastly, addressing potential security vulnerabilities in quantum cryptography systems, such as quantum hacking, and developing robust countermeasures will be essential to ensure the security of digital communication in the quantum era.

## VIII. Conclusion

This paper has highlighted the significant potential of quantum cryptography in revolutionizing the security and privacy of digital communication in the quantum era. However, numerous challenges and open research questions must be addressed to fully harness this potential. Focused research on the development of robust QKD protocols, secure post-quantum cryptographic algorithms, and efficient solutions for IoT devices is essential to enable secure and practical quantum cryptography applications, including quantum cryptocurrency. Additionally, experimental demonstrations and real-world deployments will play a crucial role in validating and refining the proposed protocols and algorithms. Through continued research and collaboration, it is anticipated that these challenges can be overcome, leading to enhanced security and privacy in digital communication and fostering the widespread adoption of quantum cryptography solutions in various domains. The insights and research directions presented in this paper aim to guide future work in this exciting and rapidly evolving field, ultimately contributing to a new era of secure communication in the quantum age.